\documentclass[twocolumn,showpacs,showkeys,pdflatex]{revtex4}
\usepackage{hyperref}
\usepackage{amsmath}
\usepackage{bm}
\usepackage{graphicx}

\begin{document}

\mark{{Delayed-coupled logistic...}	{A.C. Mart\'{\i}, M. Ponce and C. Masoller}}

\title{Dynamics of delayed-coupled chaotic logistic maps: influence of network topology, connectivity and delay times}

\author{Arturo C. Mart\'{\i}$^1$}	\email{marti@fisica.edu.uy}
\author{Marcelo Ponce C.$^1$}	\email{mponce@fisica.edu.uy}
\author{Cristina Masoller$^2$}	\email{cristina.masoller@upc.edu}

\affiliation{$^1$Instituto de F\'{\i}sica, Facultad de Ciencias, Universidad de la Rep\'ublica, Igu\'a 4225, Montevideo 11400, Uruguay}
\affiliation{$^2$Departament de Fisica i Enginyeria Nuclear, Universitat Politecnica de Catalunya, Colom 11, E-08222 Terrassa, Barcelona, Spain}


\begin{abstract}
We review our recent work on the synchronization of a
network of delay-coupled maps, focusing on the interplay of the
network topology and the delay times that take into account the finite
velocity of propagation of interactions. We assume that the elements
of the network are identical ($N$ logistic maps in the regime where
the individual maps, without coupling, evolve in a chaotic orbit) and
that the coupling strengths are uniform throughout the network.  We
show that if the delay times are sufficiently heterogeneous, for
adequate coupling strength the network synchronizes in a spatially
homogeneous steady-state, which is unstable for the individual maps
without coupling. This synchronization behavior is referred to as
``suppression of chaos by random delays'' and is in contrast with the
synchronization when all the interaction delay times are homogeneous,
because with homogeneous delays the network synchronizes in a state
where the elements display in-phase time-periodic or chaotic
oscillations. We analyze the influence of the network topology
considering four different types of networks: two regular (a ring-type
and a ring-type with a central node) and two random (free-scale
Barabasi-Albert and small-world Newman-Watts). We find that when the
delay times are sufficiently heterogeneous the synchronization
behavior is largely independent of the network topology but depends on
the networks connectivity, i.e., on the average number of neighbors
per node.
\end{abstract}

\maketitle

\section{Introduction}
A system composed of many nonlinear interacting units often forms a
complex system with new emergent properties that are not held by the
individual units. Such systems describe a wide variety of phenomena in
biology, physics, and chemistry. The emergent property is usually
synchronous oscillations. Examples include the synchronized activity
in pacemaker heart cells, the cicardian rhythms, the flashing
on-and-off in unison of populations of fireflies, synchronized
oscillations in laser arrays, in Josephson junction arrays etc.
\cite{kuramoto,stefano,pikovsky,zanette}.

The effect of time-delayed interactions, which arise from a realistic
consideration of finite communication times, is a key issue that has
received considerable attention. The first systematic investigation of
time-delayed coupling was done by Schuster and Wagner
\cite{schuster_PTP_1989}, who studied two coupled phase oscillators
and found multistability of synchronized solutions.  Since then,
delayed interactions have been studied in the context of linear
systems \cite{jirsa_PRL_2004}, phase oscillators \cite{phase_osc},
limit-cycle oscillators \cite{limit_cycl}, coupled maps
\cite{maps,atay_PRL_2004}, lasers \cite{jordi,mandel}, neurons
\cite{foss,neuronal,misha}, etc.

It is well-known that oscillators that interact with different delay
times can synchronize \cite{otsuka,longtin,zanette_PRE_2000,
atay_distributed_delays,marti,epl_2007}, but the interplay of the
interaction delays and the network topology is still poorly
understood. Here we review the main results of our recent work on the
synchronization of chaotic logistic maps
\cite{nosotros_prl,nosotros_pre,nosotros_physicaA} focusing on the
interplay of delays and topology. We show that a network of
delay-coupled logistic maps can synchronize, for adequate coupling
strength, in spite of the fact that the interactions among the maps
have an heterogeneous distribution of delay times. In the synchronized
state the chaotic dynamics of the individual maps is suppressed and
all elements of the network are in a steady-state, which is an
unstable fixed-point of the uncoupled maps. This is in sharp contrast
with the synchronized dynamics when the delays are homogeneous
(instantaneous coupling and fixed-delay coupling), because with
uniform delays the maps evolve in either periodic or chaotic orbits.

We investigate the influence of the network topology considering four
different types of networks: two regular (a ring-type and a ring-type
with a central node) and two random (free-scale Barabasi-Albert
\cite{barabasi-RMP} and small-world Newman-Watts) \cite{nw_1999}. We
find that steady-state synchronization depends mainly on the average
number of neighbors per node but is largely independent of the network
architecture (i.e., the way the links are distributed among the
nodes). This is also in contrast with the homogeneous delay case,
because when the delay times are uniform the synchronization of the
network depends strongly on the connection topology
\cite{atay_PRL_2004}.

The rest of the paper is structured as follows. In Sec.~\ref{sec:mod}
we introduce the basic ingredients of our model.  Subsequently,
Sec.~\ref{sec:res} presents the results of the simulations.  Finally,
Sec.~\ref{sec:dis} summarizes the main findings and presents the
conclusions from this investigation.

\section{Networks and Interactions Model}	\label{sec:mod}
We consider a network of $N$ coupled maps:
\begin{equation}
\label{mapa} x_i(t+1)= (1-\epsilon) f[x_i(t)] + \frac{\epsilon}
{b_i} \sum_{j=1}^N \eta_{ij} f[x_j(t-\tau_{ij})],
\end{equation}
where $t$ is a discrete time index, $i$ is a discrete spatial index
($i=1\dots N$), $f(x)=ax(1-x)$ is the logistic map, the matrix
$\eta=(\eta_{ij})$ defines the connectivity of the network:
$\eta_{ij}=\eta_{ji}=1$ if there is a link between the $i$th and $j$th
nodes, and zero otherwise. $\epsilon$ is the coupling strength, which
is uniform throughout the network, and $\tau_{ij}$ is the delay time
in the interaction between the $i$th and $j$th nodes (the delay times
$\tau_{ij}$ and $\tau_{ji}$ need not be equal). The sum in
Eq.~(\ref{mapa}) runs over the $b_i$ nodes which are coupled to the
$i$th node ($b_i = \sum_j \eta_{ij}$). The normalized pre-factor
$1/b_i$ means that each map receives the same total input from its
neighbors.

It can be noticed that the homogeneous steady-state
\begin{equation}
x_i(t)=x_j(t)=x_0, \forall i,j,t,
\end{equation}
where $x_0$ is a fixed point of the uncoupled map, $x_0=f(x_0)$, is
a solution of Eq.~(\ref{mapa}) regardless of the delay times and of
the connectivity of the network.

We will consider delay times that are either homogeneous
($\tau_{ij}=\tau_0$ for all $i$, $j$) or heterogeneous. In the latter
case we introduce a disorder parameter, $c$, that quantifies the
degree of heterogeneity and allows varying the delays from a delta
distribution (uniform delays) to Gaussian or exponential
distributions. Specifically we consider

i) $\tau_{ij} = \tau_0 + c \xi$, where $\xi$ is Gaussian distributed
with zero mean and standard deviation one. The delays are homogeneous
($\tau_{ij}=\tau_0$) for $c=0$ and are Gaussian distributed around
$\tau_0$ for $c\ne 0$ [depending on $\tau_0$ and $c$ the distribution
of delays has to be truncated to avoid negative delays, see
Fig. 1(f)].

ii) $\tau_{ij} = \tau_0 + c \xi$, where $\xi$ is exponentially
distributed, positive, with unit mean. The delays are homogeneous
($\tau_{ij}=\tau_0$) for $c=0$ and are exponentially distributed,
decaying from $\tau_0$ for $c\ne 0$.

To investigate the influence of the topology we consider four
networks, two of them are regular and the other two are random. The
regular ones are a ring of nearest-neighbor elements (NN) while in the
second one we added a central node connected to all other nodes
(ST). The random networks consist of a scale free network (SF)
constructed according to the Barabasi-Albert method
\cite{barabasi-RMP} and, concerning the last one, we use the
small-world (SW) topology proposed by Newman and Watts \cite{nw_1999}.

In the next section we present results of simulations that show that
the homogeneous state-state, Eq.~(2), with $x_0$ being the nontrivial
fixed point of the logistic map, $x_0=1-1/a$, is a stable solution for
adequate coupling strength and delay times.

\section{Results}	\label{sec:res}
The simulations were done choosing a random initial configuration,
$x_i(0)$ randomly distributed in [0,1], and letting the network evolve
initially without coupling [in the time interval
$0<t<\max(\tau_{ij})$]. We present results for $a=4$, corresponding to
fully developed chaos of the individual maps, but similar results have
been found for other values of the parameter $a$.

With both, either heterogeneous or homogeneous delay times, if the
coupling strength is large enough the network synchronizes in a
spatially homogeneous state: $x_i=x_j$ $\forall$ $i,j$. Figures 1 and
2 display the transition to synchronization as $\epsilon$
increases. In these figures the network has a small-world topology
\cite{nw_1999}, but similar results have been found for other
topologies, as discussed below. At each value of $\epsilon$, 100
iterates of an element of the network are plotted after transients.
To do these bifurcation diagrams only the coupling strength,
$\epsilon$, was varied; the network connectivity, $\eta_{ij}$, the
delay times, $\tau_{ij}$, and the initial conditions, $x_i(0)$, are
the same for all values of $\epsilon$.

\begin{figure}
\includegraphics[width=.8\columnwidth]{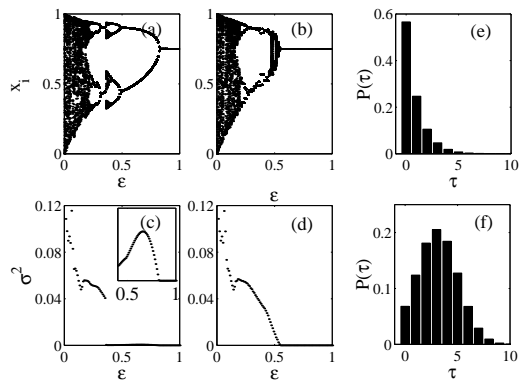}
\caption{$x_i$ vs. $\epsilon$ (a) and (b); $\sigma^2$ vs. $\epsilon$,
(c) and (d).  In Figs. 1(a) and 1(c) the delays are distributed
exponentially ($\tau_0=0$, $c=1.2$, see text); the distribution is
shown in Fig. 1(e).  In Figs. 1(b) and 1(d) the delays are Gaussian
distributed ($\tau_{0}=3$, $c=2$, see text); the distribution is shown
in Fig. 1(f). The inset in Fig. 1(d) shows with detail the transition
to synchronization: $\sigma^2$ decreases abruptly at $\epsilon \sim
0.4$, and is zero for $\epsilon > 0.8$.  $N=500$, $a=4$, and $p=0.3$.}
\end{figure}

\begin{figure}
\centerline{
\includegraphics[width=.8\columnwidth]{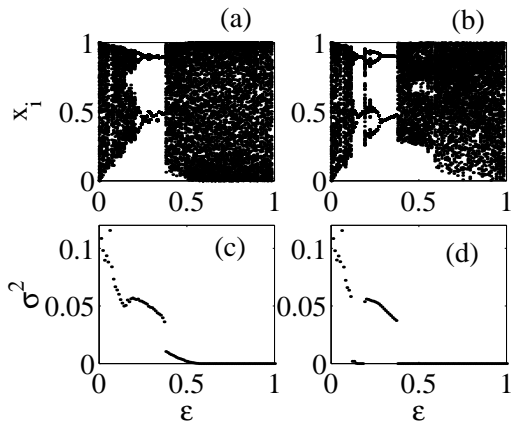}}
\caption{$x_i$ vs. $\epsilon$ (a) and (b); $\sigma^2$ vs. $\epsilon$
(c) and (d). In (a),(c) $\tau_{ij}=0$ $\forall$ $i$,$j$; in (b),(d)
$\tau_{ij}=3$ $\forall$ $i$,$j$. All other parameters as in Fig. 1.}
\end{figure}

Figure 1(a) displays results for exponentially distributed delays and
Fig. 1(b) for Gaussian distributed delays; Fig. 2(a) for instantaneous
interactions, and Fig. 2(b) for uniform delays. It can be noticed that
for small $\epsilon$ the four bifurcation diagrams are very similar
(we refer to this region as the ``weak coupling region''); however, as
$\epsilon$ increases above $\sim 0.1$ the bifurcation diagrams begin
showing some differences, and for large $\epsilon$ they differ
drastically: $x_i$ is constant in Figs. 1(a) and 1(b),
$x_i=x_0=1-1/a$, while $x_i$ varies within [0,1] in Figs. 2(a) and
2(b).

To characterize the transition to synchronization we use the indicator
\begin{equation}
\sigma^2 = 1/N<\sum_i [x_i(t)-<x>]^2>_t,
\end{equation}
where $<.>$ denotes an average over the elements of the network and
$<.>_t$ denotes an average over time. Figures 1(c), 1(d), 2(c) and
2(d) display $\sigma^2$ vs. $\epsilon$ corresponding to the
bifurcation diagrams discussed above. It can be noticed that for large
$\epsilon$ there is in-phase synchronization in the four cases
[$\sigma^2=0$ if and only if $x_i(t)=x_j(t)$ $\forall$ $i$,$j$];
however, we remark that an inspection of the time-dependent dynamics
reveals that the synchronized dynamics is different: for heterogeneous
delays the maps are in a steady-state, while for homogeneous delays
the maps evolve either periodically or chaotically. It can also be
observed that the four plots of $\sigma^2$ vs. $\epsilon$ are similar
in the weak coupling regime (in Fig. 2(d) the network synchronizes
also in a window of small $\epsilon$; this occurs for odd delays as
reported in \cite{atay_PRL_2004}).

\begin{figure}[t]
\begin{center}
\includegraphics[angle=270,width=.48\columnwidth]{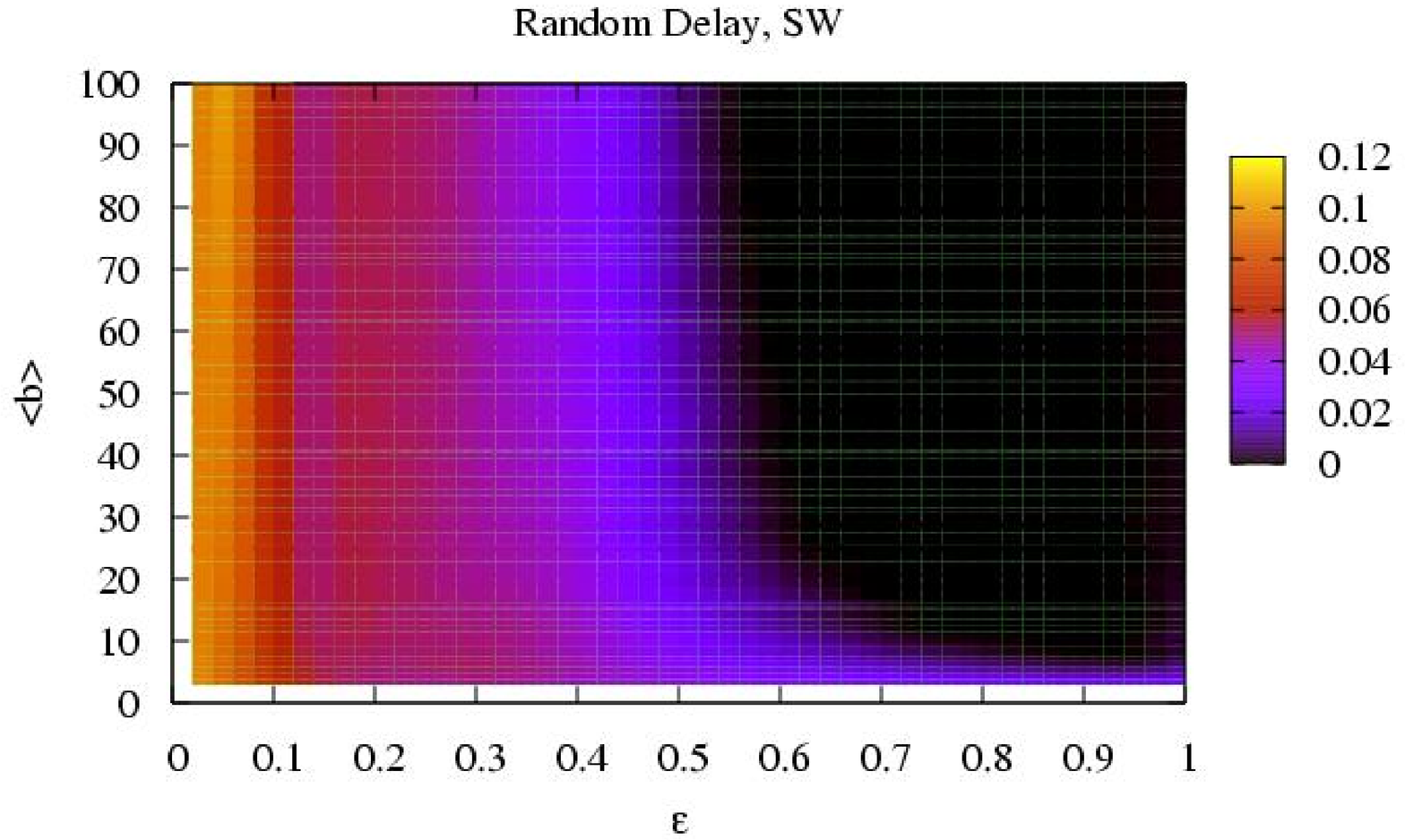}
\includegraphics[angle=270,width=.48\columnwidth]{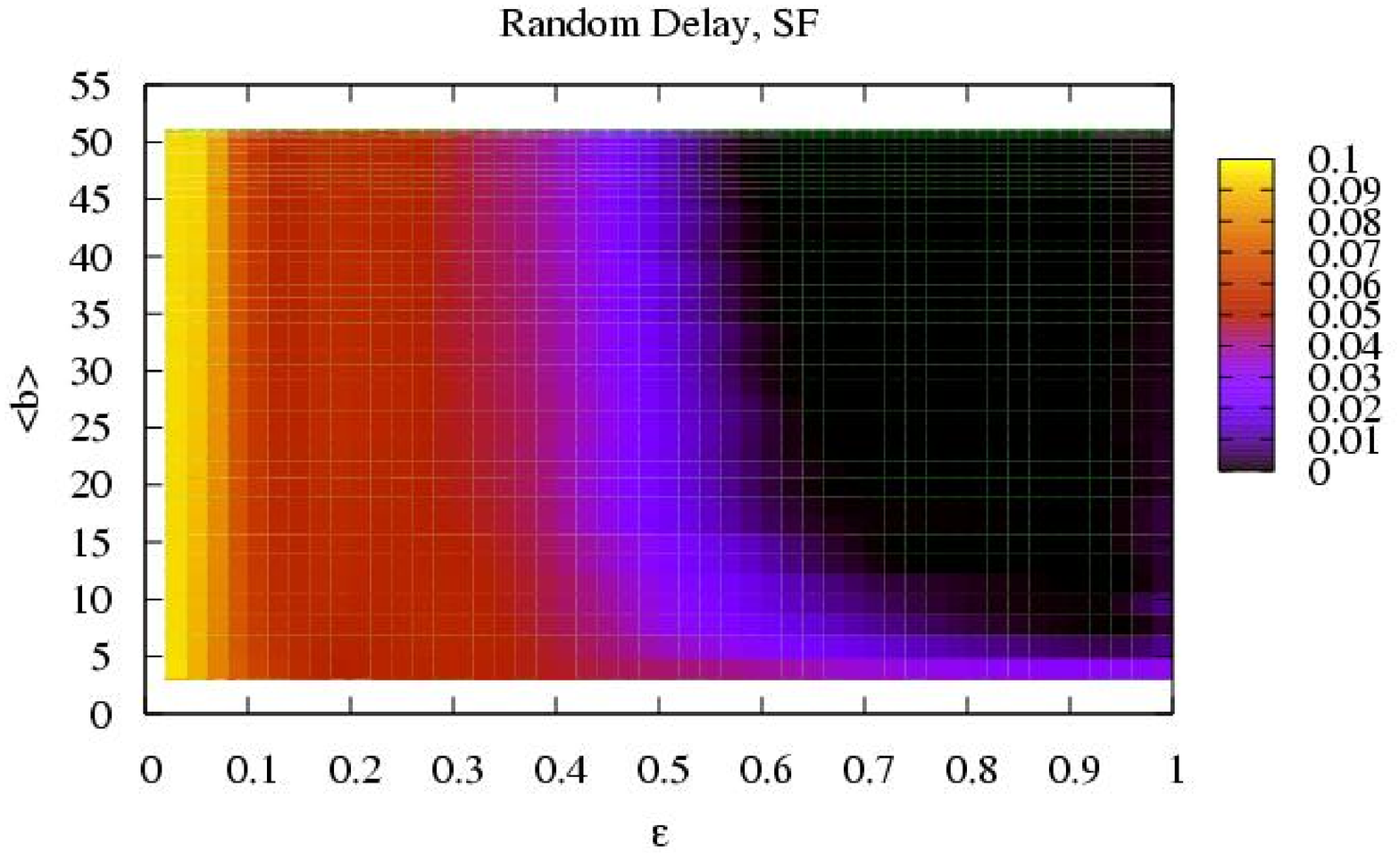}
\includegraphics[angle=270,width=.48\columnwidth]{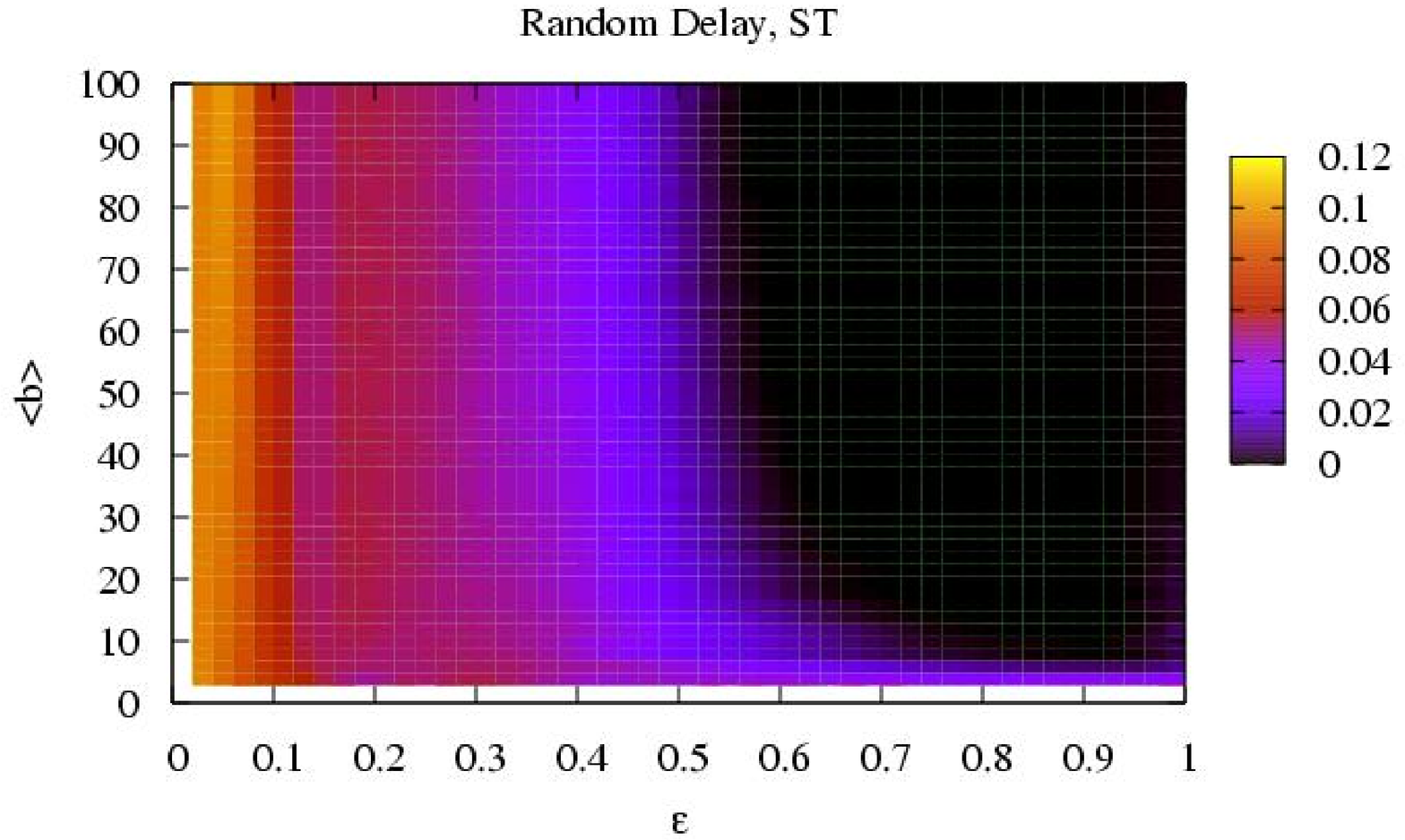}
\includegraphics[angle=270,width=.48\columnwidth]{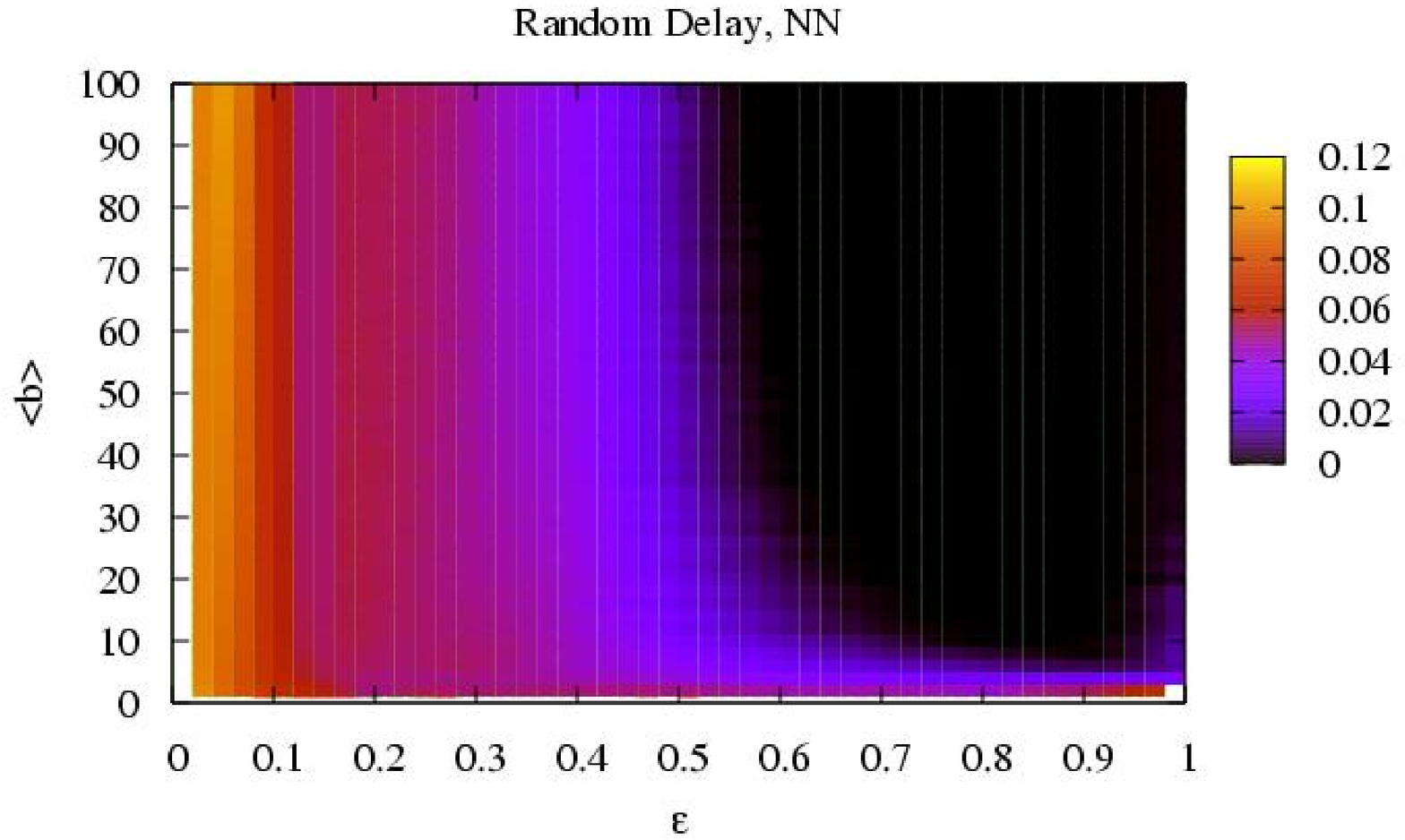}
\end{center}
\caption{(Color online) Random delays: synchronization regions for the
four different networks considered and Gaussian distributed delays.
The density plots represent the
parameter $\sigma^2$ as a function of $\epsilon$ and $b$ averaged over
10 realizations of the initial conditions ($N=100$ and $a=4$). Up: SW and
SF, down: ST and NN.}
\label{fig-rd}
\end{figure}

Next we analyze the influence of the network
topology. Figure~\ref{fig-rd} displays a density plot of $\sigma^2$ as
a function of the coupling strength, $\epsilon$, and the mean number
of neighbors per node, $<b>=\sum b_i/N$. The four panels correspond to
the different networks mentioned above (SW, SF, ST and NN); in these
cases the Gaussian distribution of delays is the same in the four
panels. Despite the differences when the number of neighbors is small,
it can be observed that the synchronizability of the network is
largely independent of the topology.

\begin{figure}
\begin{center}
\includegraphics[angle=270,width=.48\columnwidth]{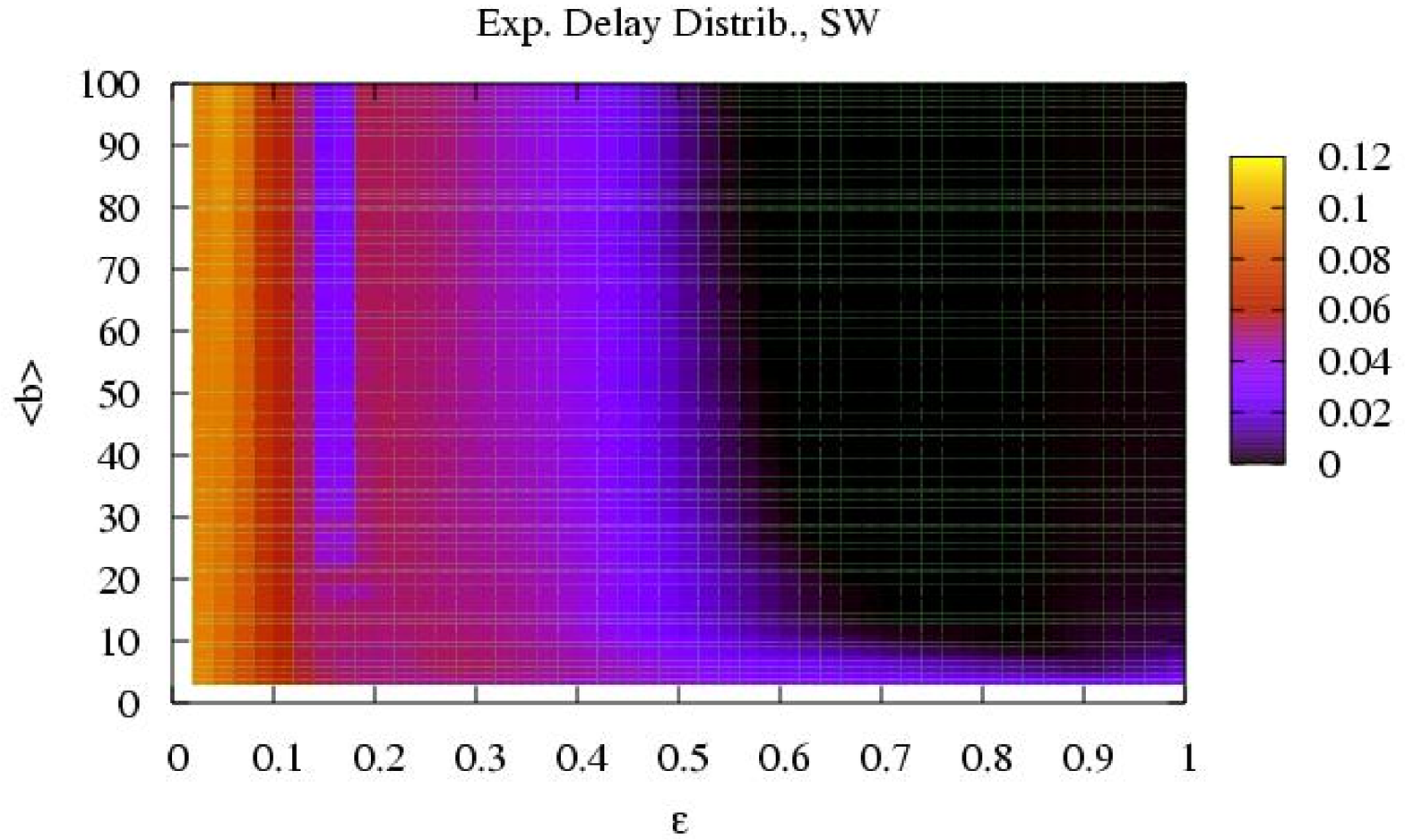}
\includegraphics[angle=270,width=.48\columnwidth]{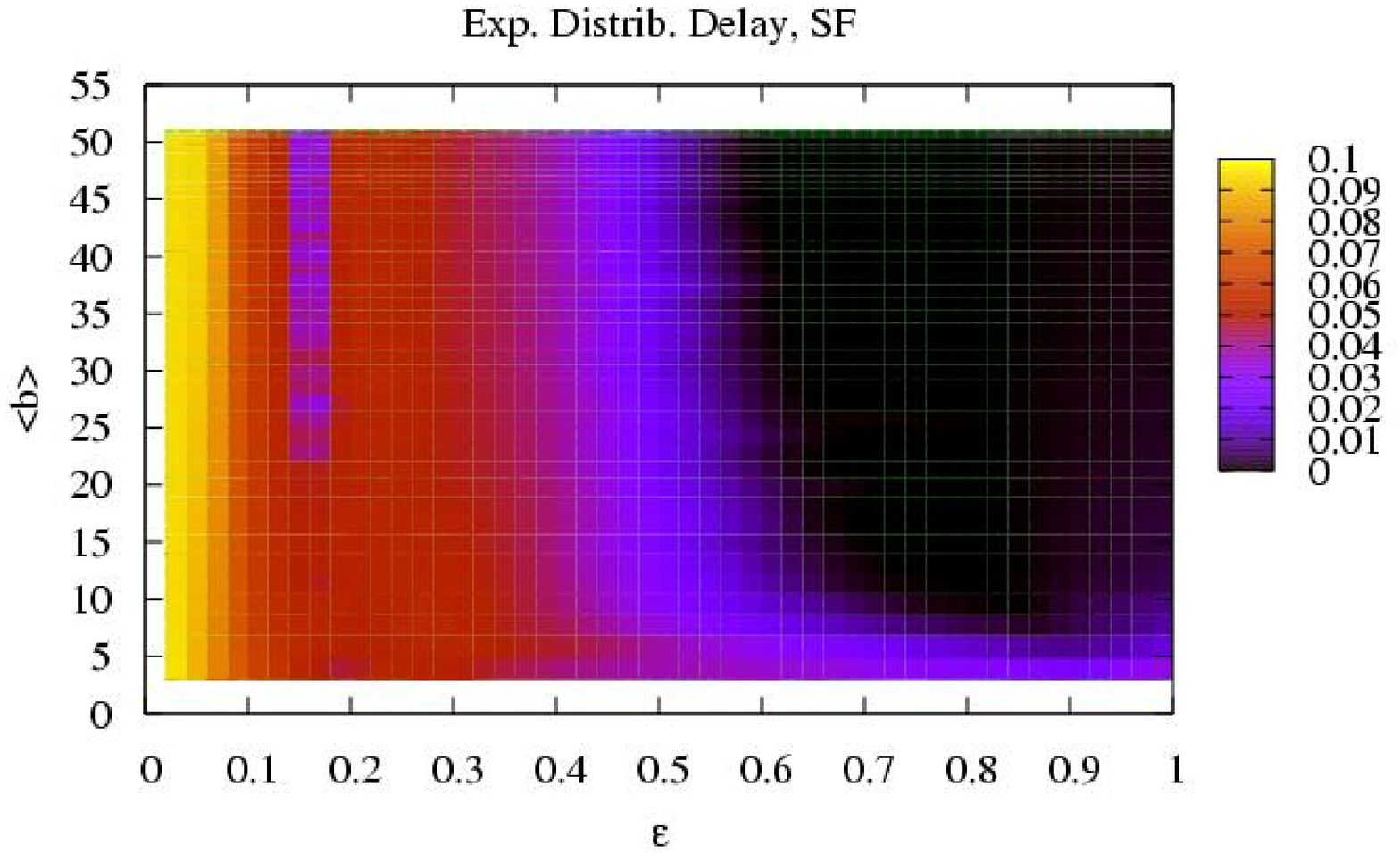}
\includegraphics[angle=270,width=.48\columnwidth]{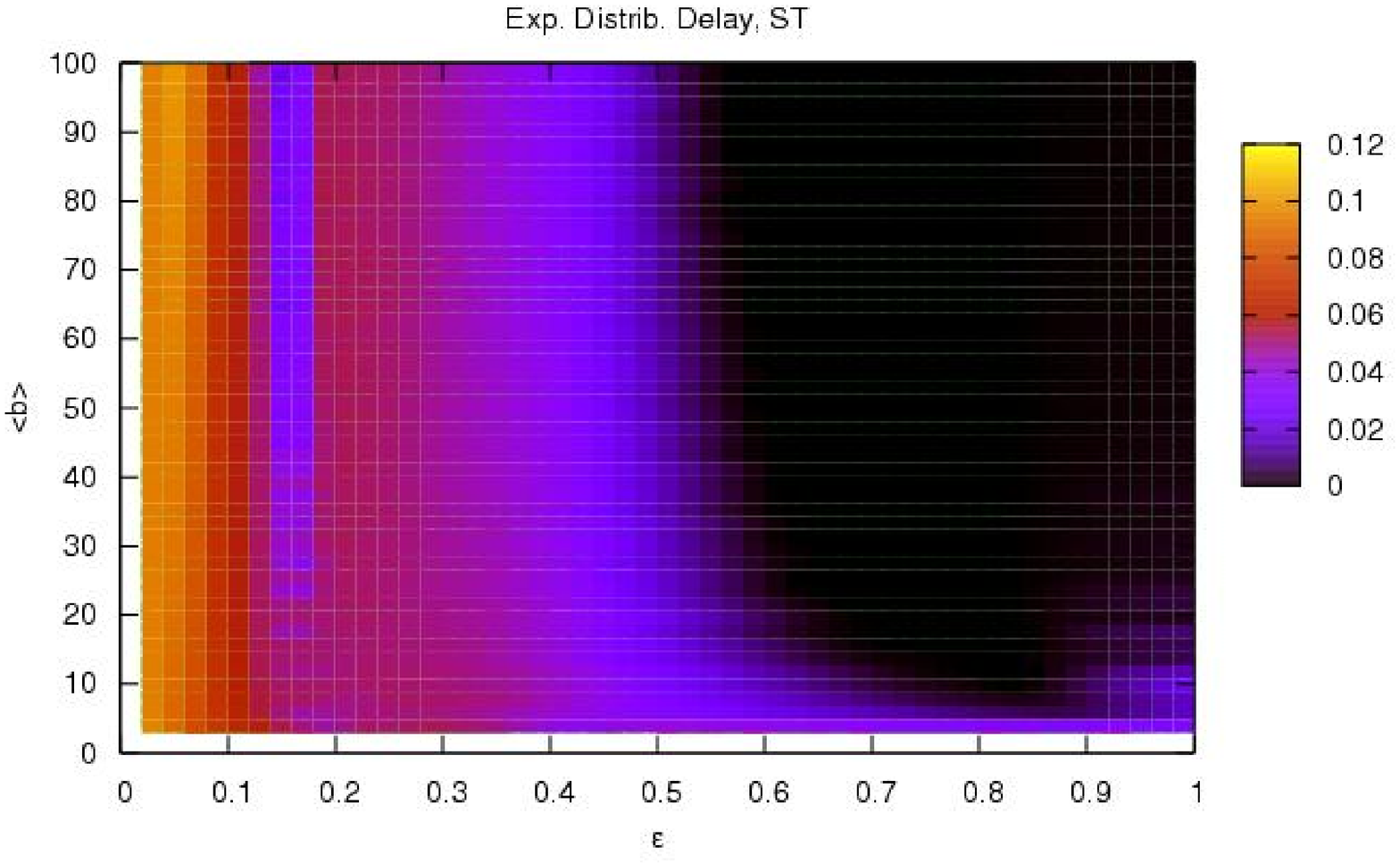}
\includegraphics[angle=270,width=.48\columnwidth]{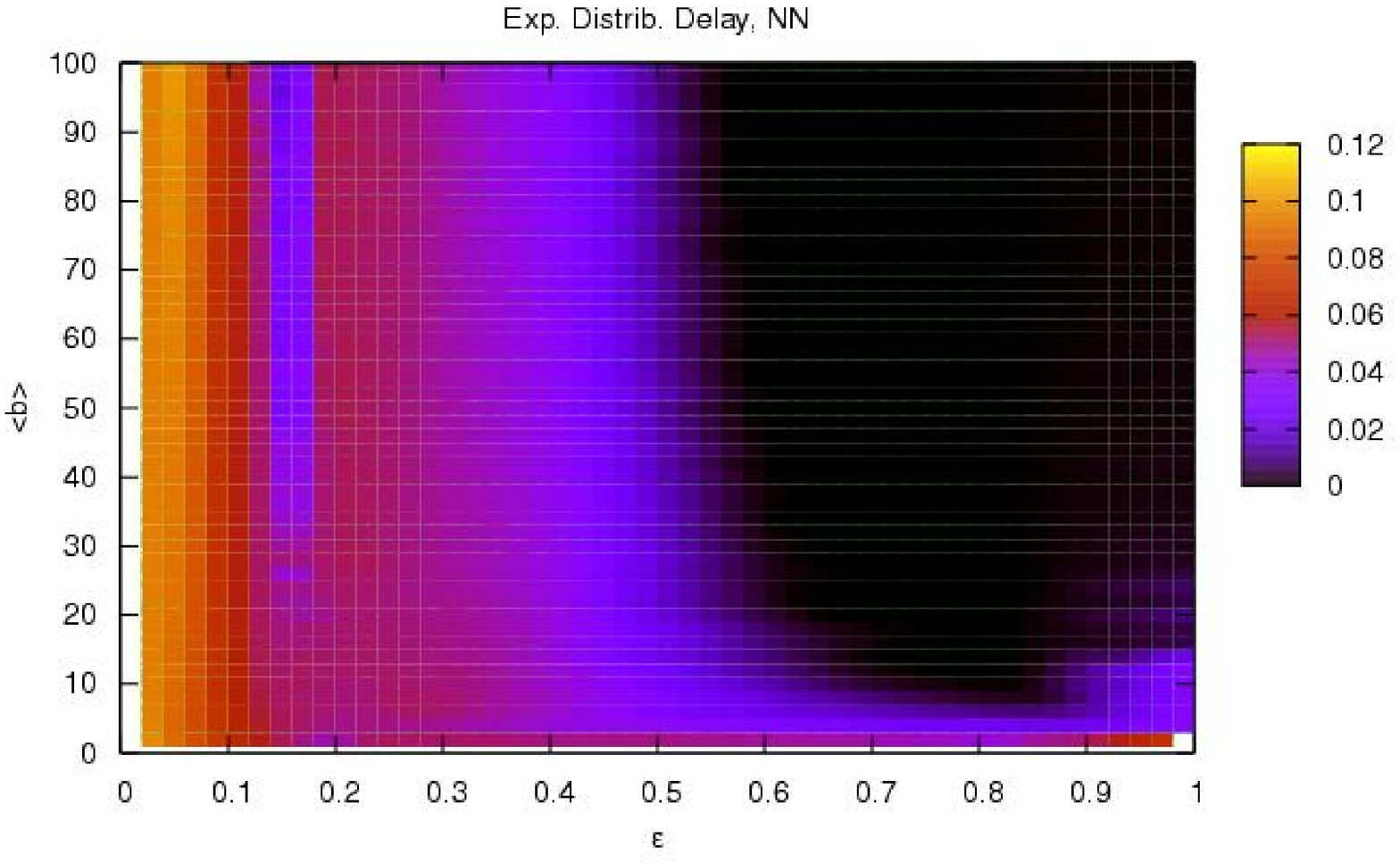}
\end{center}
\caption{(Color online) Random delays: synchronization regions for
the SW, SF, ST and NN networks and exponential distributed delays.
Same parameters as Fig.~\ref{fig-rd}. }
\label{fig-rd_exp}
\end{figure}

We also show in Fig.~\ref{fig-rd_exp} for the sake of comparison two topologies (SW, NN)
with a Gaussian delays distribution. 
Analyzing Fig.~\ref{fig-rd_exp} it is possible to conjecture that it represents
an intermediate behaviour between the Random Delay Gaussian distribution and the Fixed Delay interaction.
Regarding the case of odd delays, it looks like to remain a trace of the ``island of synchronization'' 
present in the fixed delays interactions, but which is not present in the Gaussian Random Delay distribution.

\begin{figure}[hb]
\begin{center}
\includegraphics[angle=270,width=.48\columnwidth]{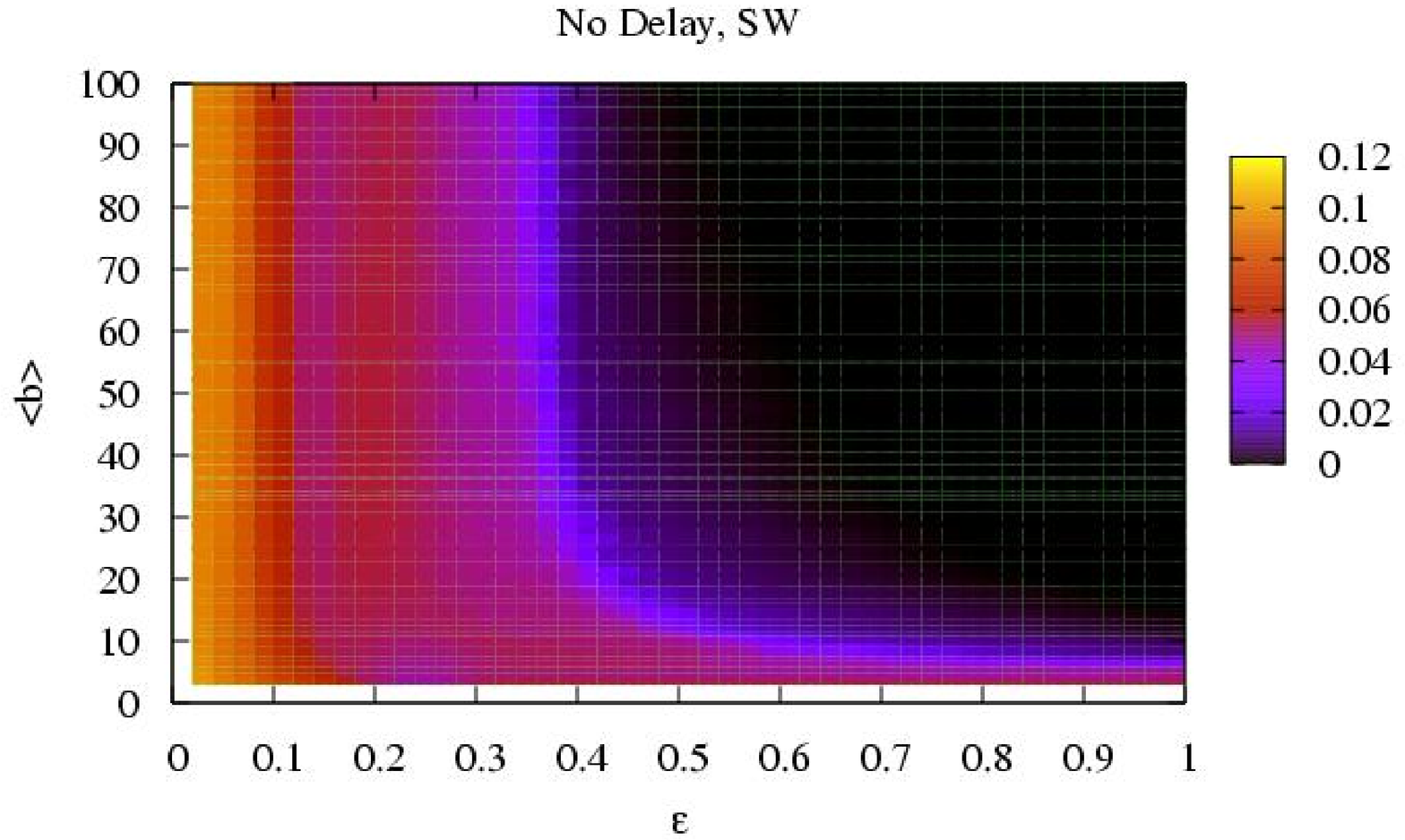}
\includegraphics[angle=270,width=.48\columnwidth]{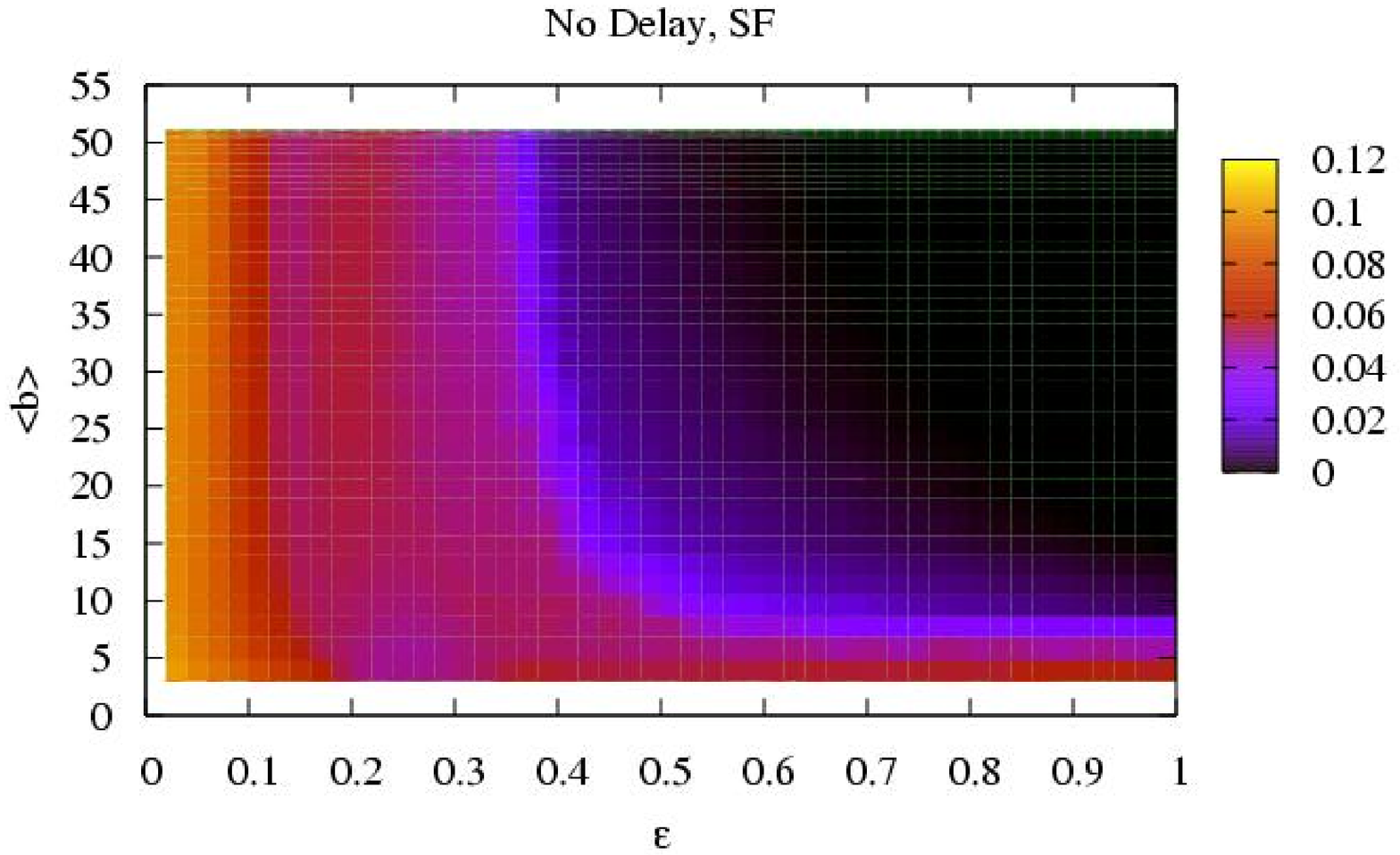}
\includegraphics[angle=270,width=.48\columnwidth]{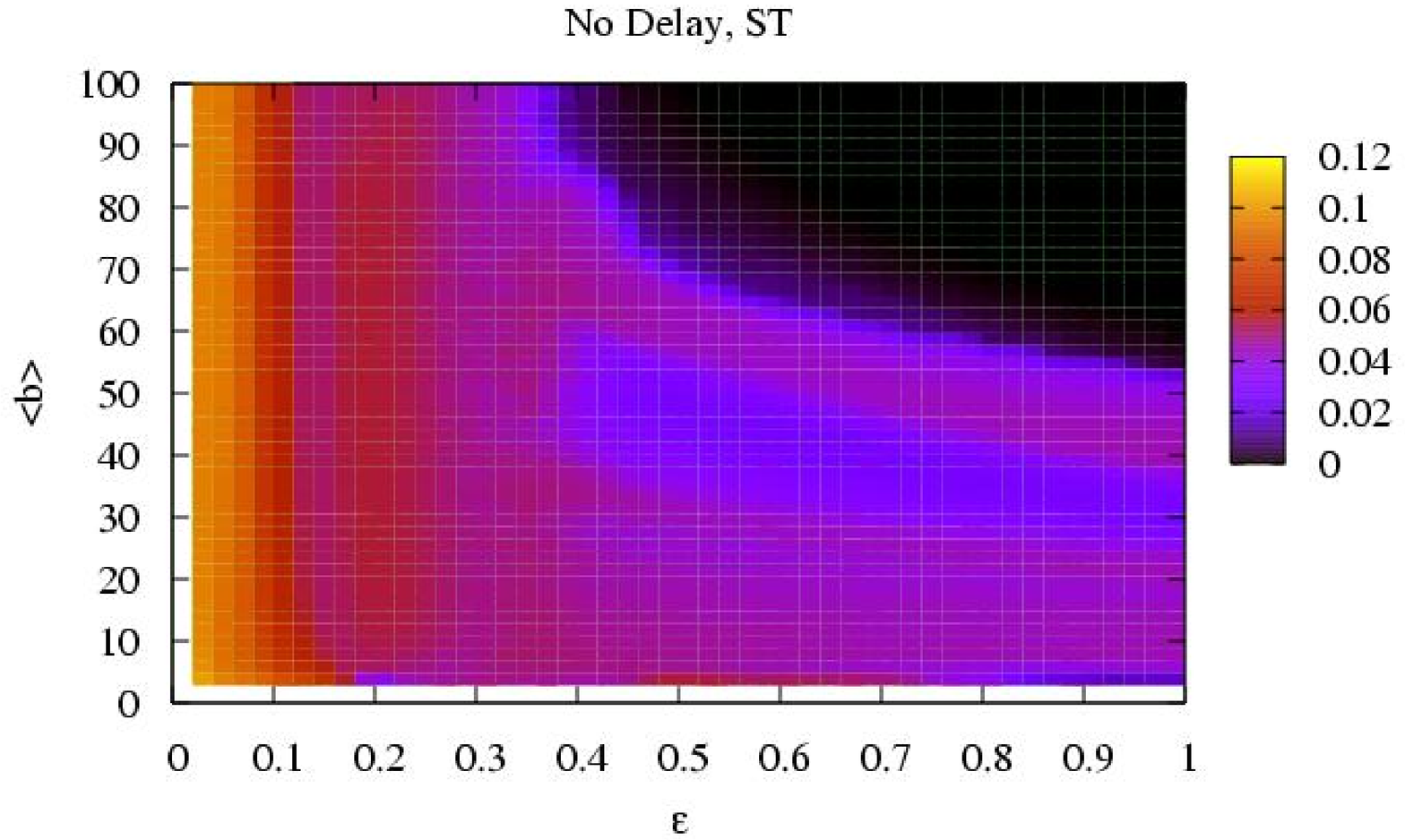}
\includegraphics[angle=270,width=.48\columnwidth]{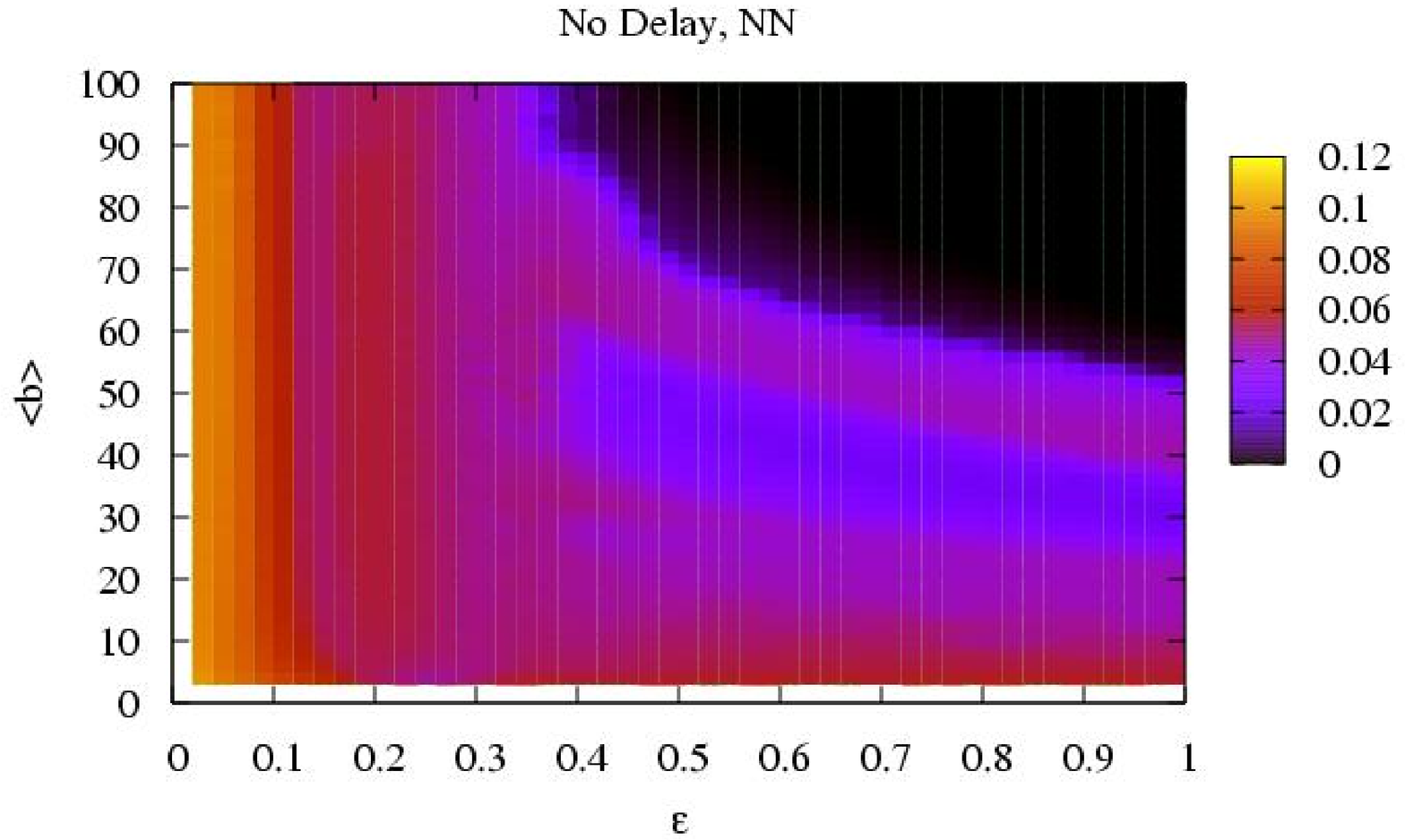}
\end{center}
\caption{(Color online) Instantaneous interactions: synchronization
regions for the four different networks considered. Same parameters as
Fig.~\ref{fig-rd}. Up: SW and SF, down: ST and NN.}
\label{fig-nd}
\end{figure}

\begin{figure}
\begin{center}
\includegraphics[angle=270,width=.48\columnwidth]{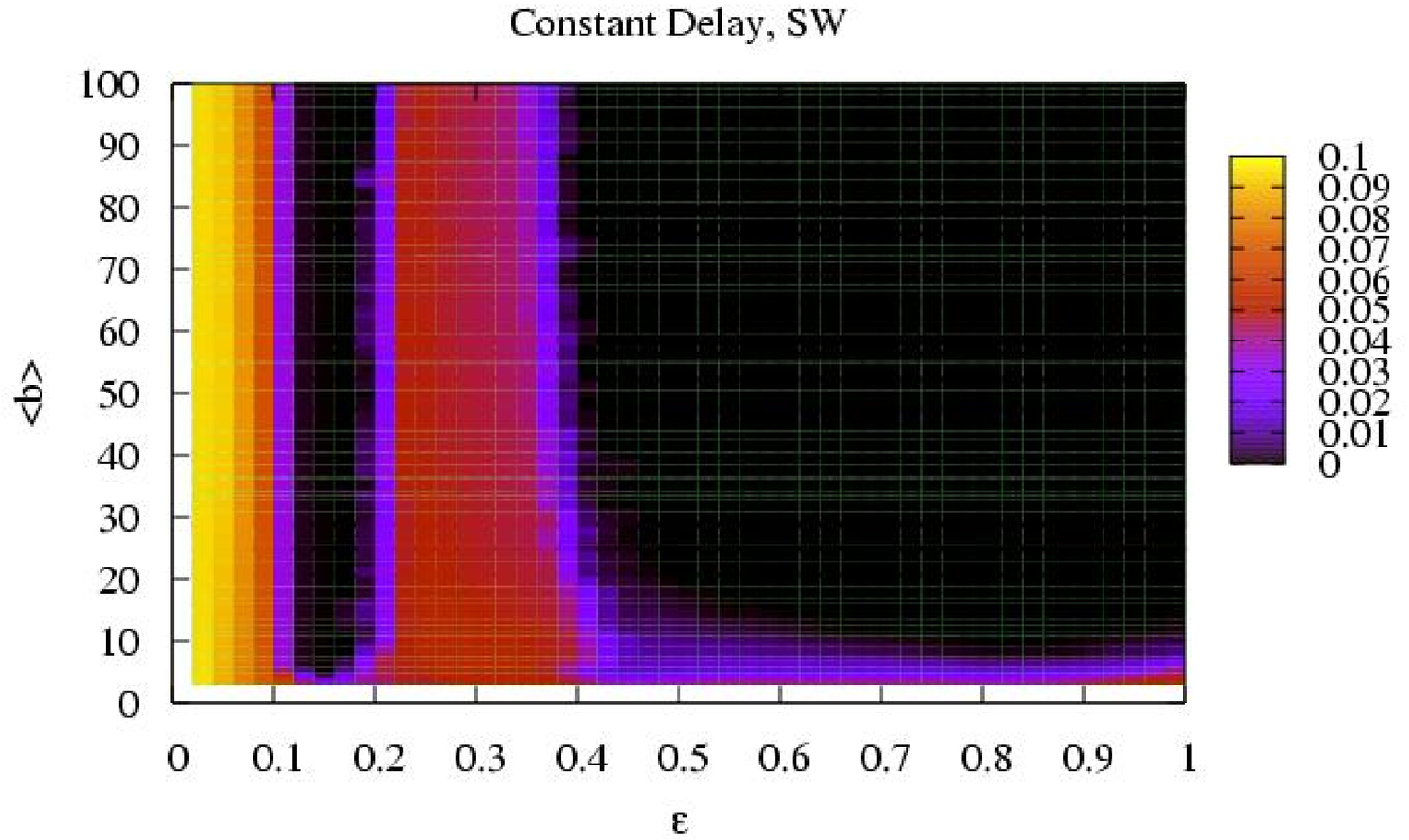}
\includegraphics[angle=270,width=.48\columnwidth]{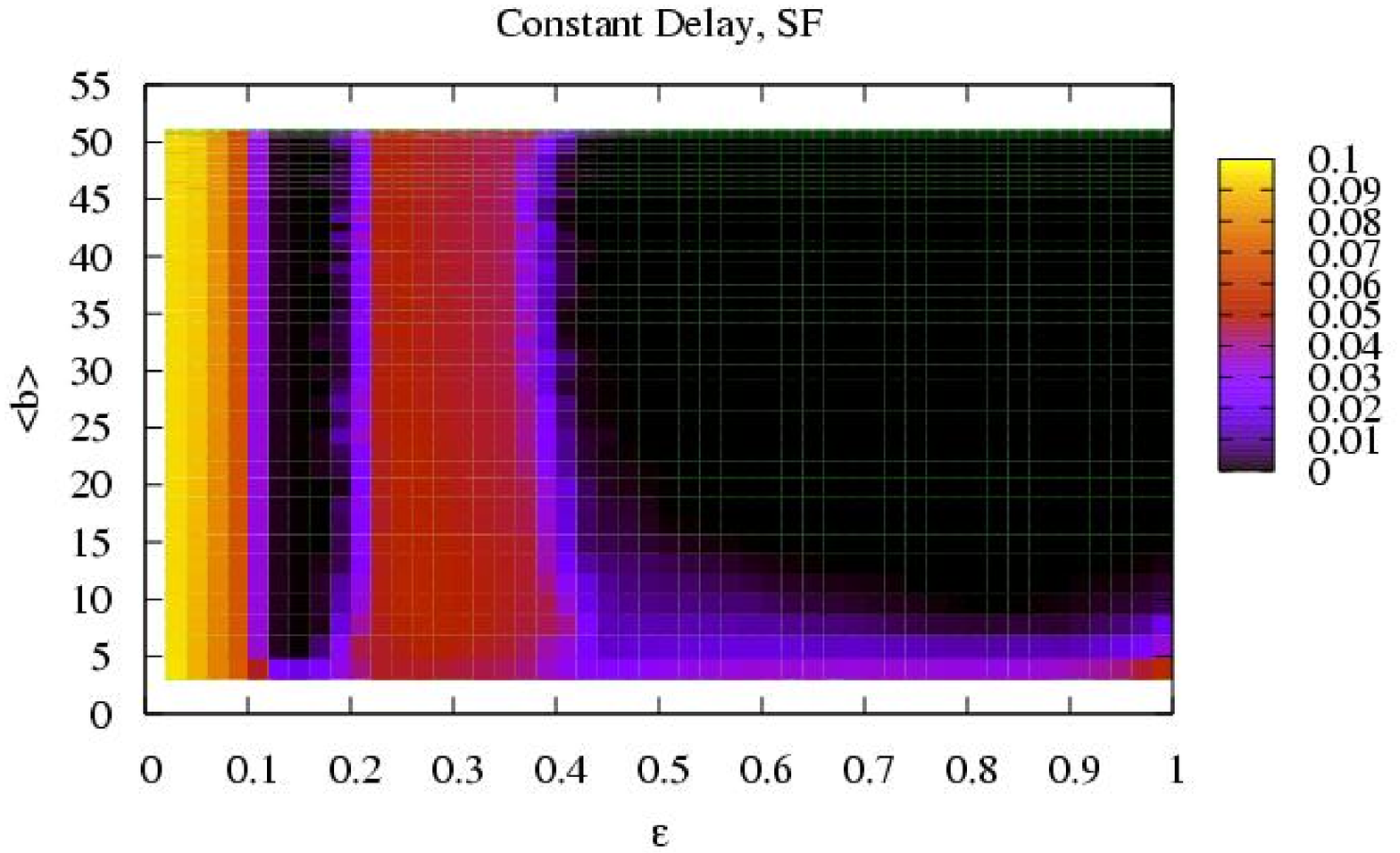}
\includegraphics[angle=270,width=.48\columnwidth]{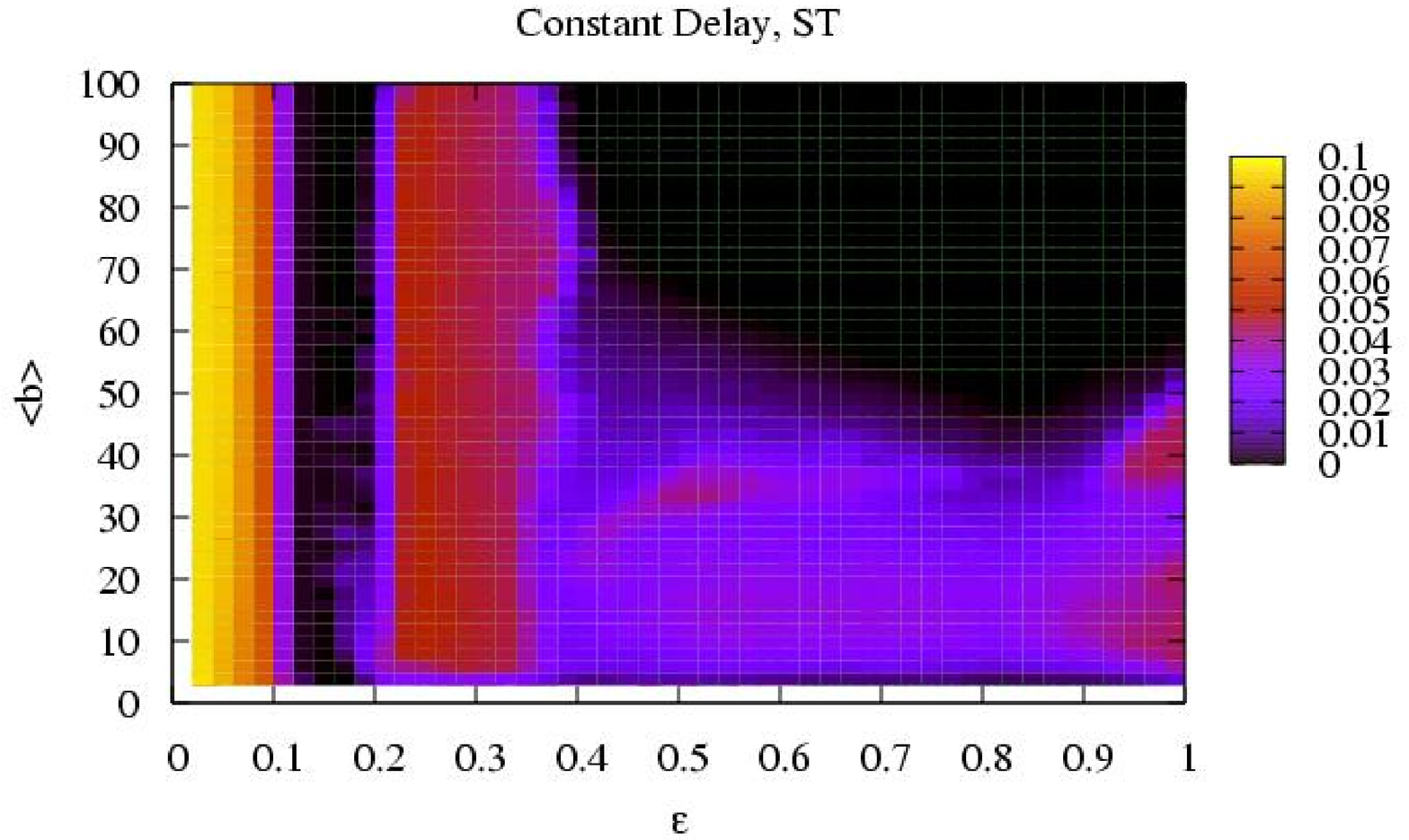}
\includegraphics[angle=270,width=.48\columnwidth]{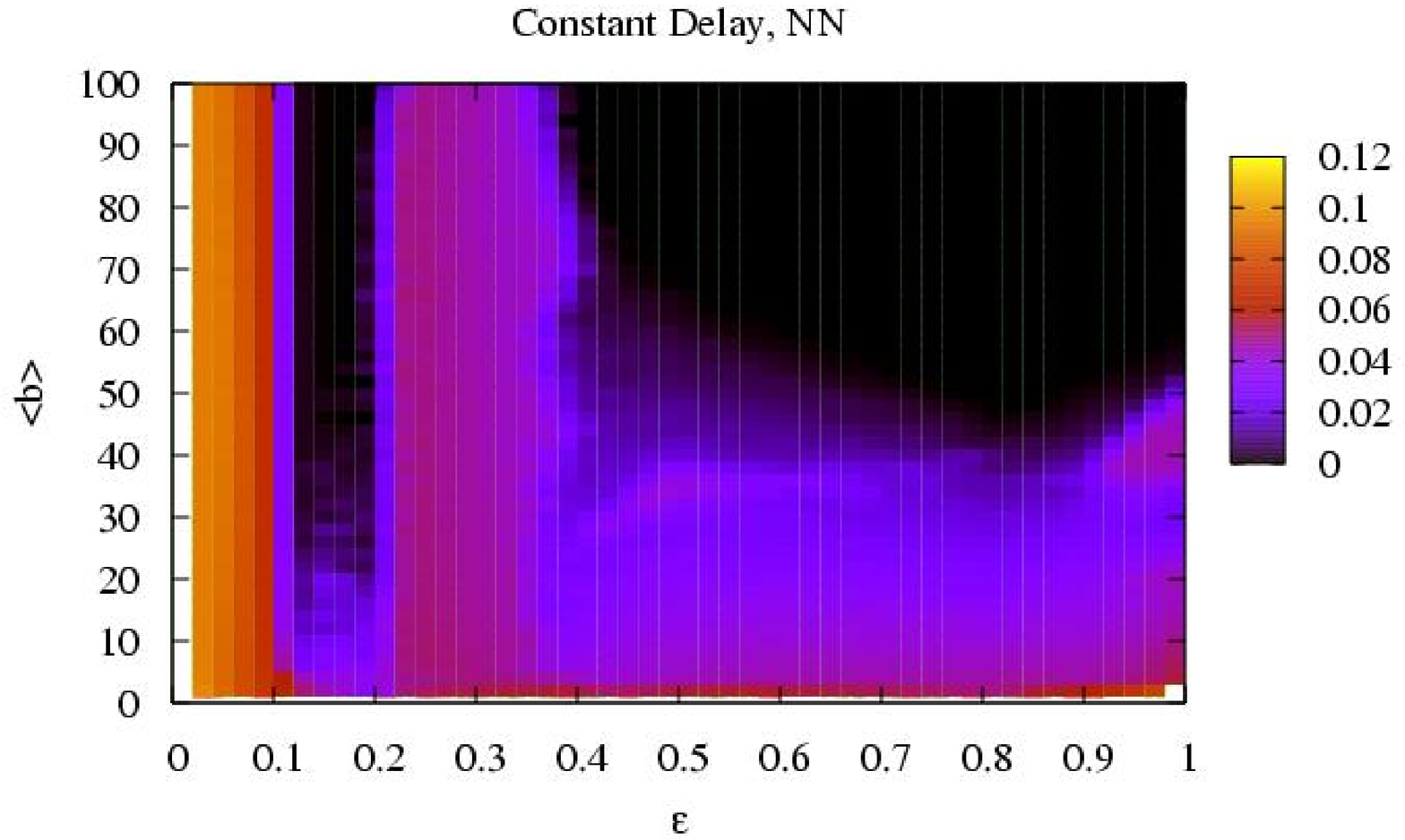}
\end{center}
\caption{(Color online) Homogeneous delays ($\tau_0=5$):
synchronization regions for the four different networks
considered. Same parameters as Fig.~\ref{fig-rd}.}
\label{fig-fd_impar}
\end{figure}

\begin{figure}
\begin{center}
\includegraphics[angle=270,width=.48\columnwidth]{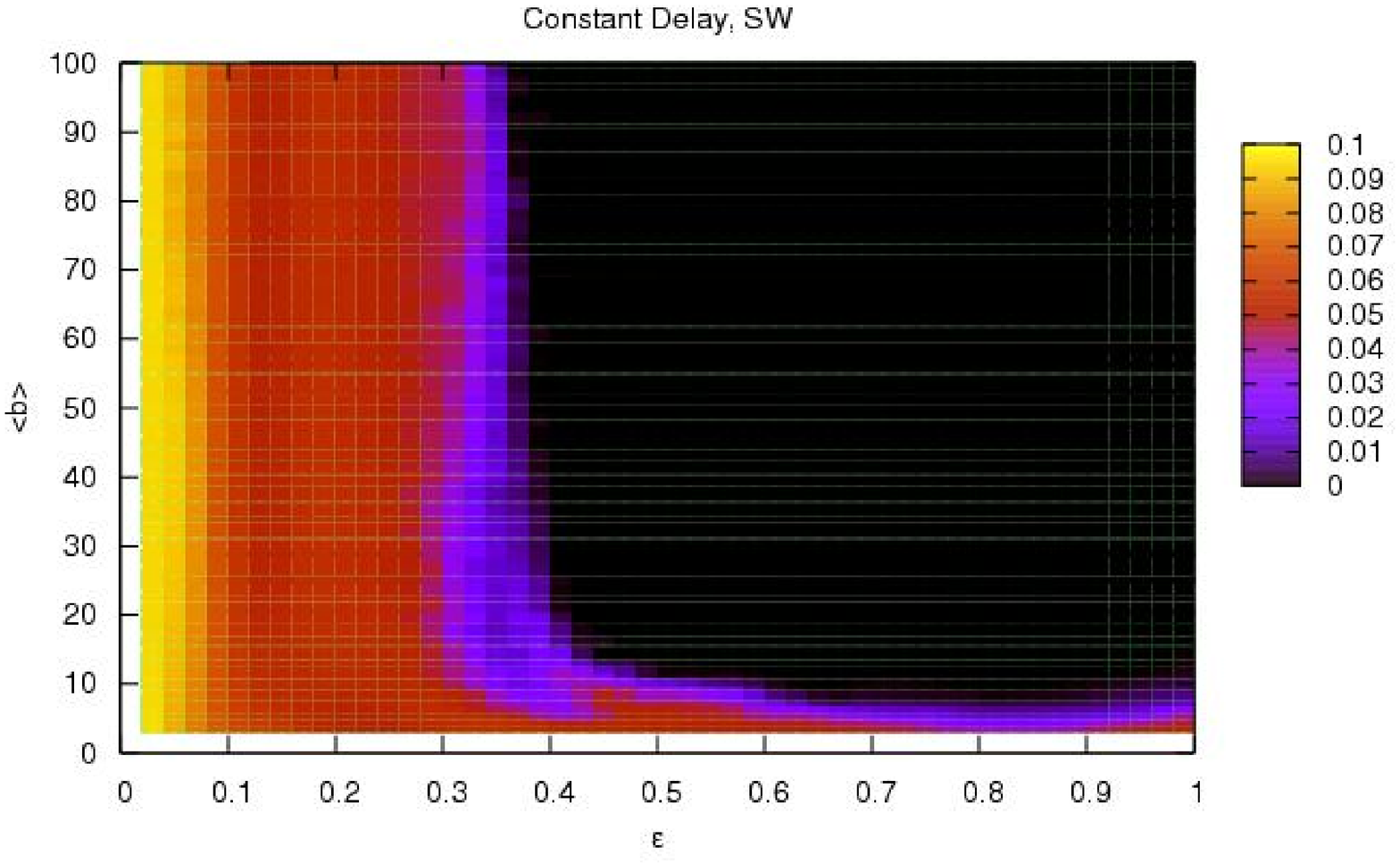}
\includegraphics[angle=270,width=.48\columnwidth]{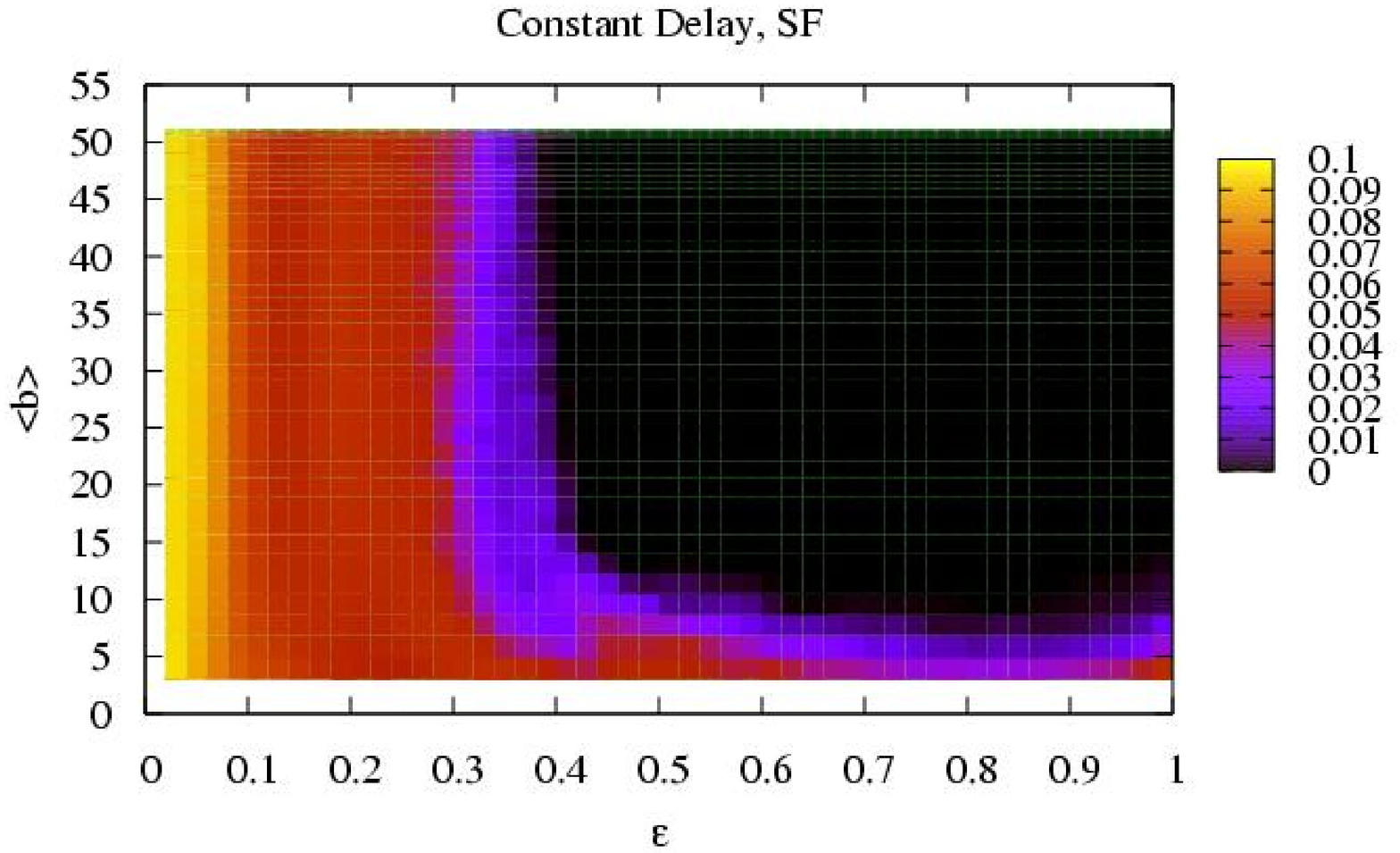}
\includegraphics[angle=270,width=.48\columnwidth]{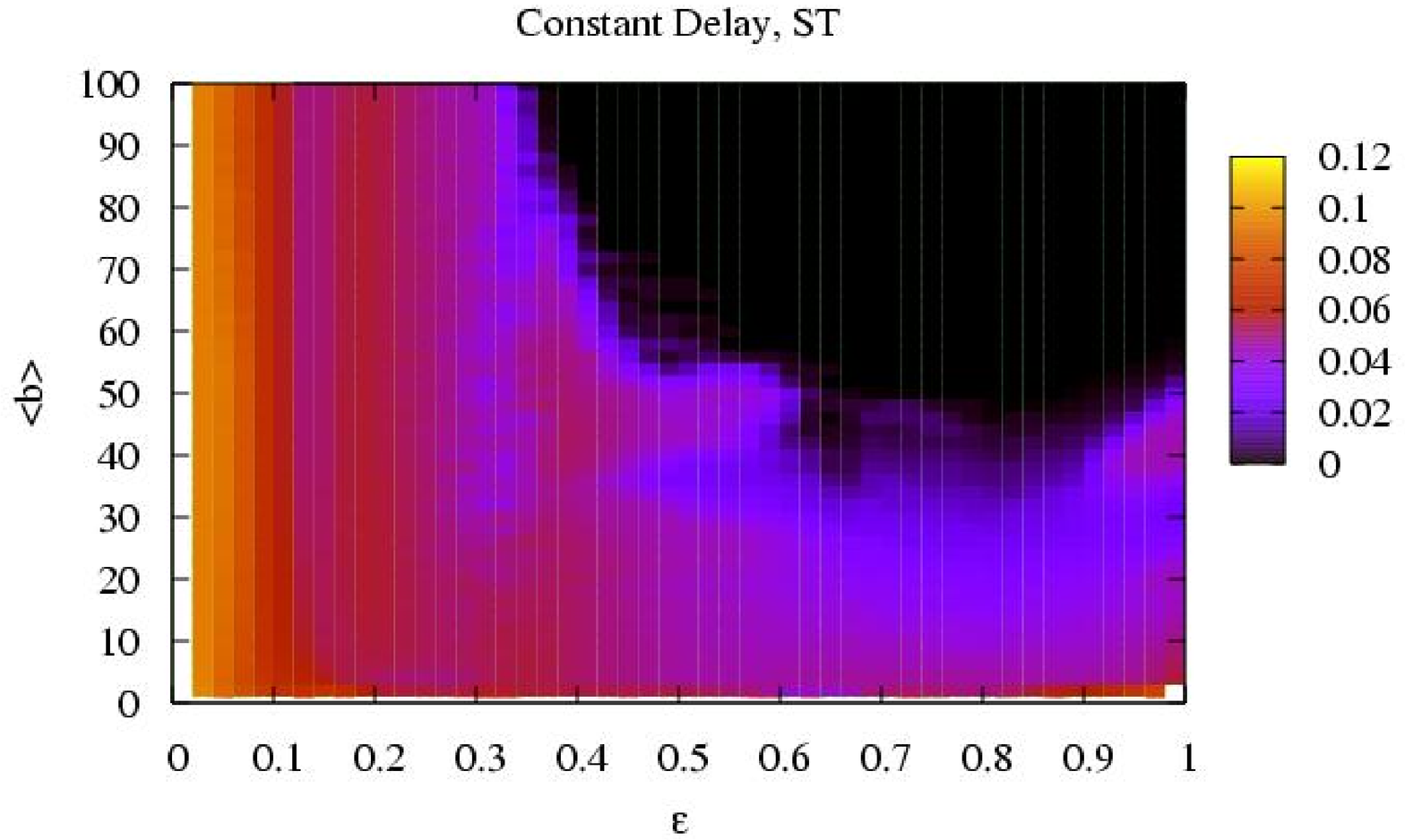}
\includegraphics[angle=270,width=.48\columnwidth]{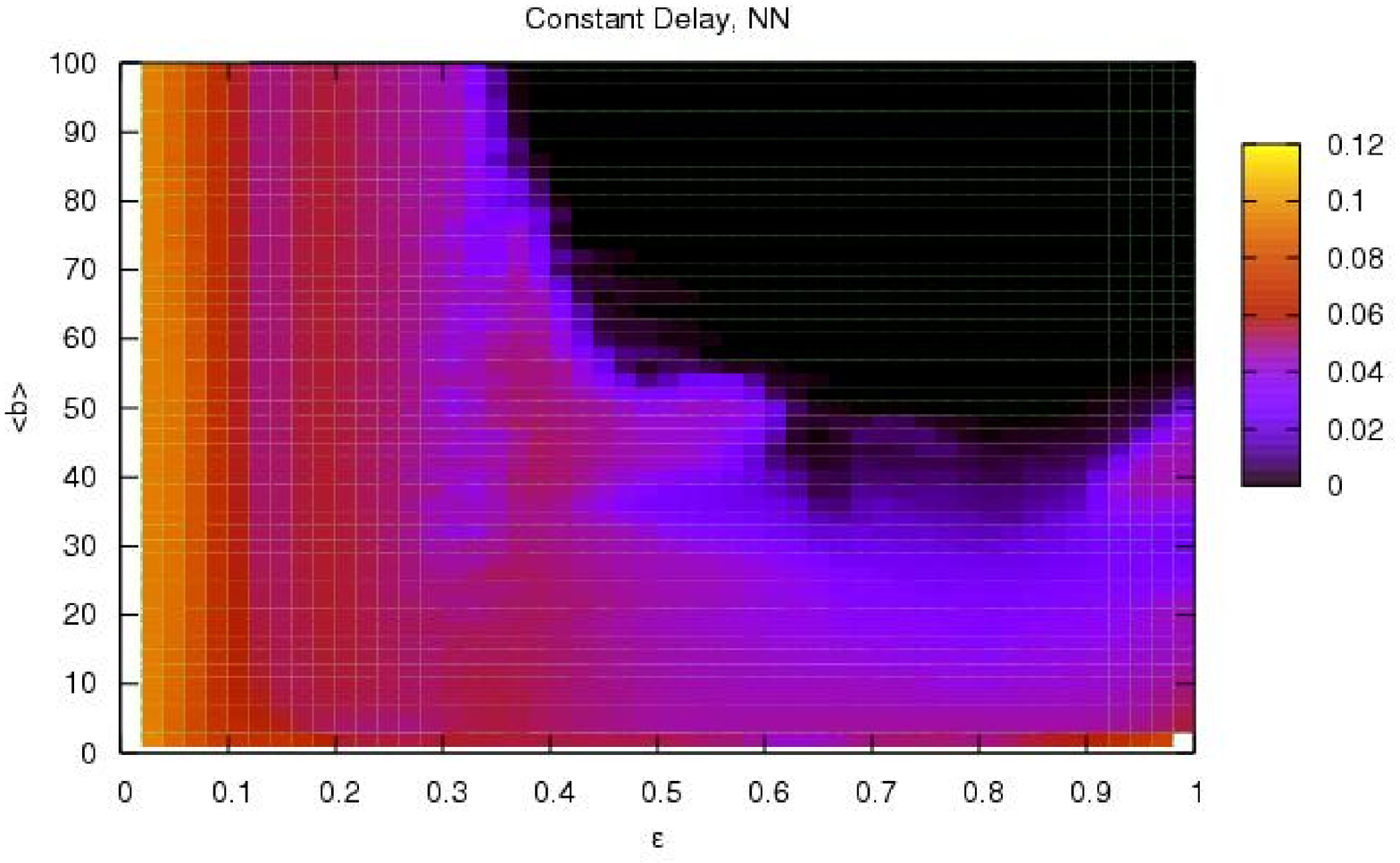}
\end{center}
\caption{(Color online) Homogeneous delays ($\tau_0=4$): synchronization
regions for the SW (up-left), SF (up-right), ST (down-left) and NN
(down-right) networks. Same parameters as Fig.~\ref{fig-rd}.}
\label{fig-fd_par}
\end{figure}

Concerning the influence of the delay distribution, in
Fig.~\ref{fig-nd} and Figs.~\ref{fig-fd_impar}-\ref{fig-fd_par} we
show the synchronization regions for the cases of instantaneous
interactions and homogeneous delays respectively. It has been reported
\cite{atay_PRL_2004} that there is a very important difference between
the case with an even or odd homogeneous delay
(Figs.~\ref{fig-fd_impar}-\ref{fig-fd_par}).  With an even delay it
can be prove that there is a kind of ``island of synchronization'', it
is a region for $\epsilon $ (roughly speaking from $\epsilon \sim$
0.15 to 0.19 -weak coupling regime-) where the maps synchronize. In
addition to this, we see a subcritical period doubling bifurcation in
this region. For an odd delay, this synchronization region disappears.
Although this kind of delay, in the case of even delays, secure these
``islands of synchronization'' there are some particularities to take
into account.  For example, in the case of the NN topology, for small
number of neighbors the synchronization is not as good as in the rest
of the topologies. Also in the upper limit of the {\it weak coupling
regime} ($\epsilon \gtrsim 0.4$) there are differences between the
irregular-networks (SW, SF) and the regular ones (NN, ST).

Considering all the showed delay distributions
(Figs.~\ref{fig-rd},\ref{fig-rd_exp},\ref{fig-nd},\ref{fig-fd_impar},\ref{fig-fd_par})
it is possible to conclude that the random delay
distributions homogenizes the synchronization propensity in relation
to the network topologies, and it also sets the value of the coupling
constant $\epsilon$ for the on-set of synchronization in a greater
$\epsilon$ compared with the other delay distributions.
For instance, for random delays a ``strong-coupling regime'' starts
roughly in $\epsilon \sim 0.55$, while for homogeneous interactions this
regime starts before $\epsilon \sim 0.4$, what is more for instantaneous
interactions (no delay at all) it is not clear determined these region,
i.e. because it depends on the topology and the average number of
neighbours.

In further researches we plan to extend this investigation about the
special weak-coupling regime and the properties of synchronization
related to the network topology and connectivity.  There we have
analyzed the properties briefly discussed here, considering five
different kind of topologies \cite{CAM_2007}.

\section{Discussion and conclusions}	\label{sec:dis}
To summarize, we studied the dynamics of a network of chaotic maps
with delayed interactions and showed that the network synchronizes in
a spatially homogeneous steady-state if the delay times are
sufficiently heterogeneous. This behavior resembles the so-called
``amplitude (or oscillator) death'' phenomenon, which refers to the fact
that under certain conditions the amplitude of delay-coupled
oscillators shrinks to zero \cite{limit_cycl}. It has been shown that
that distributed delays are more stabilizing than fixed delays
\cite{atay_distributed_delays}. The stabilization of the fixed-point
is also related to the multiple delay feedback method proposed by
Ahlborn and Parlitz \cite{parlitz}, for stabilizing unstable steady
states. We have recently compared the dynamics of a map of a network
of $N$ delayed-coupled maps with the dynamics of a map with $N$
self-feedback delayed loops \cite{nosotros_pre}. If $N$ is
sufficiently large, we found that the dynamics of a map of the network
is similar to the dynamics of a map with self-feedback loops with the
same delay times. Several delayed loops stabilize the fixed point;
however, the distribution of delays played a key role: if the delays
are all odd a periodic orbit (and not the fixed point) was stabilized.

\begin{acknowledgments}
We acknowledge financial support from PEDECIBA and CSIC
(Uruguay). A.C.M.  thanks the Abdus Salam ICTP for the hospitality and
support on occasion of the workshop Perspectives on Nonlinear Dynamics
2007. CM acknowledges partial support from the European
Commission (GABA project, FP6-NEST 043309).

\end{acknowledgments}

\end{document}